\documentclass{article}
\usepackage{spconf,amsmath,graphicx,hyperref}
\usepackage{booktabs}
\usepackage{multicol}
\usepackage{multirow}
\usepackage{graphicx}

\hypersetup{hidelinks}

\title{Spike Sorting with VanillaSort}
\name{
Zishuo Feng$^{1,2}$ and
Feng Cao$^{2}$\sthanks{Corresponding author: {f.cao@students.uu.nl}.}
}

\address{
$^{1}$University Medical Center Utrecht, The Netherlands\\
$^{2}$Utrecht University, The Netherlands\\
{z.feng5@students.uu.nl}, {f.cao@students.uu.nl}
}
\begin{document}
%
\maketitle
\begin{abstract}
Training spike detectors on real recordings is challenging because
algorithmically generated labels can be noisy and incomplete.
We propose VanillaSort, combining multichannel detection with
spatially augmented, template-guided clustering. VanillaDet uses
visibility-aware masking, truncated Gaussian targets and a
temporally tolerant positive-bag loss, followed by conditional
event-SNR gating. VanillaCluster combines HuiduRep embeddings
with relative-amplitude features for Gaussian mixture clustering
and refines assignments using cross-fitted waveform templates
built from selected core events. VanillaDet improves detection accuracy
over SimSort by two and three percentage points on the static
and drift subsets of Hybrid Janelia, respectively. The complete
pipeline also improves sorting performance over the corresponding
HuiduRep baselines. These results support learning from imperfect real-data
labels and incorporating waveform consistency into neuronal
assignment.

\end{abstract}
\section{Introduction}
\label{sec:intro}
Extracellular electrodes measure voltage fluctuations associated with nearby neuronal activity. Action potentials produce brief extracellular waveforms, known as spikes. A single spike may appear on neighboring channels, while each channel can contain activity from multiple neurons \cite{10.3389/fninf.2022.851024}. Spike sorting identifies these events and assigns them to individual neurons. In modular spike sorting pipelines, detection locates candidate events, then feature extraction and clustering support neuronal assignment. Low signal-to-noise ratios (SNRs) can obscure weak events, while electrode drift, the relative movement between the probe and surrounding tissue, can alter recorded waveforms over time \cite{pachitariu2024spike}. Reliable event-level annotations are difficult to obtain at scale, motivating training on automated sorting outputs. However, these outputs may contain false positives and missed events and must be treated as pseudo-labels rather than ground truth.



\textbf{Related Work.} Current detectors in spike sorting frameworks can be divided into two categories: classical methods and learning-based methods \cite{cao2026huidureprobustselfsupervisedframework}. Existing classical approaches can include amplitude-threshold detection \cite{quiroga2004unsupervised} and template-based detection with deconvolution, which is used by Kilosort4 \cite{pachitariu2024spike}. Learning-based methods include the neural detector in YASS \cite{lee2017yass} and the simulation-trained Transformer detector in SimSort \cite{zhang2025simsortdatadrivenframeworkspike}. Besides, HuiduRep combines contrastive learning and denoising for waveform representation, while its published pipeline uses threshold detection and Gaussian-mixture clustering \cite{cao2026huidureprobustselfsupervisedframework}.

\textbf{Our Contributions.} To address the unreliability of algorithmically generated labels and improve both spike detection and overall spike sorting performance, we propose VanillaSort. VanillaSort consists of two modules: VanillaDet for spike detection and VanillaCluster for neuronal assignment.

VanillaDet is a multichannel spike detector trained with visibility-aware uncertainty supervision. It accounts for two sources of label uncertainty: an event annotated from the full probe may not be observable within a local channel neighborhood, while an unlabeled interval does not necessarily correspond to background activity. To further improve temporal localization, VanillaDet employs Gaussian targets together with a temporally tolerant positive-bag loss \cite{dietterich1997solving}.

\begin{figure*}
    \centering
    \includegraphics[width=0.86\textwidth]
        {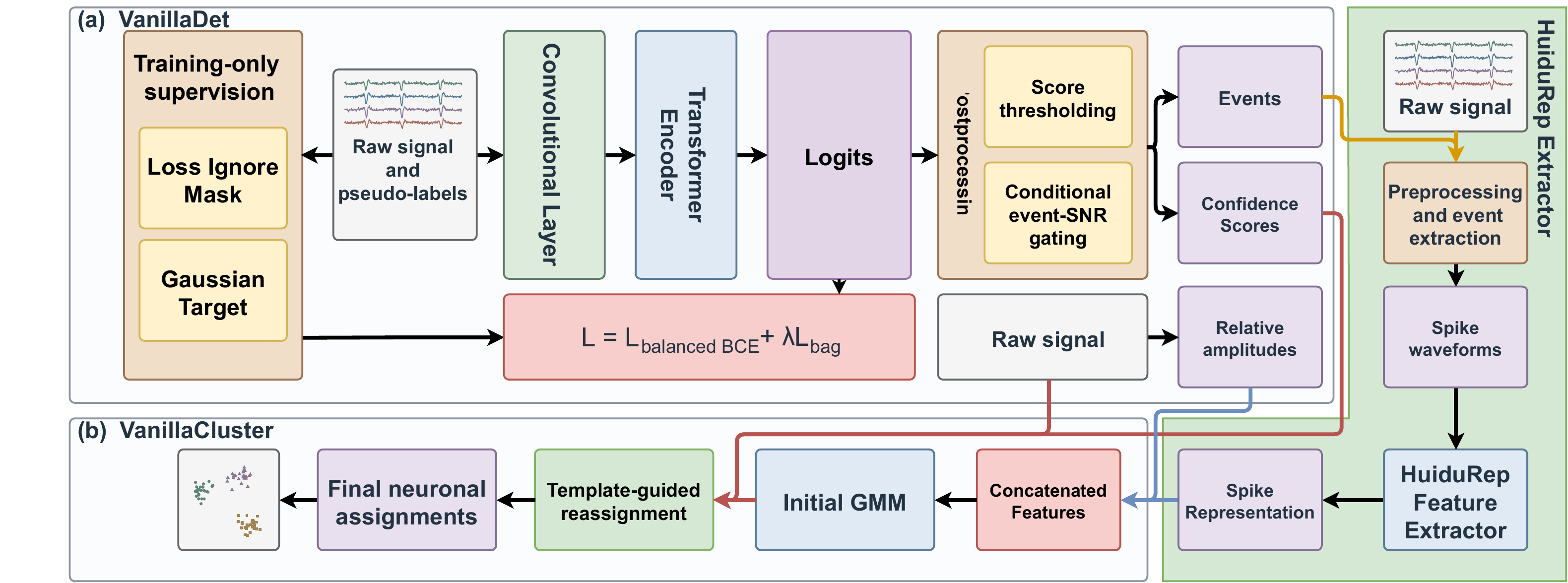}
    \caption{Overview of the VanillaSort pipeline.  
    (a) VanillaDet combines a convolutional frontend and a Transformer
    encoder for spike detection.
    (b) VanillaCluster combines HuiduRep representations with
    cross-channel relative-amplitude features for initial GMM
    clustering followed by template-guided reassignment.}
    \label{fig:vanillasort_architecture}
\end{figure*}

VanillaCluster augments HuiduRep embeddings \cite{cao2026huidureprobustselfsupervisedframework} with cross-channel relative-amplitude features for Gaussian mixture clustering (GMM) \cite{pearson1894contributions}. It then refines spike-to-unit assignments using cross-fitted waveform templates, incorporating waveform consistency while preserving the original GMM evidence.

On Hybrid Janelia datasets \cite{magland2020spikeforest}, VanillaDet outperforms SimSort \cite{zhang2025simsortdatadrivenframeworkspike}. Besides, replacing the threshold detector and the cluster in the HuiduRep pipeline also improves sorting performance. Overall improvements are larger on drift recordings than on static recordings.

\section{Methodology}
\label{sec:methodology}
As shown in Fig. \ref{fig:vanillasort_architecture}, the VanillaSort pipeline consists of three main components: VanillaDet, the HuiduRep feature extractor \cite{cao2026huidureprobustselfsupervisedframework}, and VanillaCluster.
\subsection{VanillaDet}
\begin{figure}[t]
    \centering
    \includegraphics[width=0.80\columnwidth]{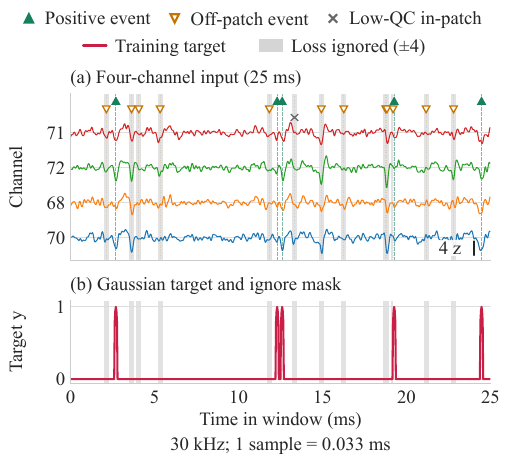}
    \caption{
        VanillaDet training supervision.
        (a) Four-channel input with different event types.
        (b) Gaussian targets and ignored regions for loss computation.
    }
    \label{fig:vanilladet_supervision}
\end{figure}

To reduce unreliable supervision from noisy pseudo-labels,
we introduce an uncertainty mask and Gaussian targets
(Fig.~\ref{fig:vanilladet_supervision}). Events annotated using
the full probe may not be clearly observable within the local
input, particularly when their primary channels lie outside
the input patch or their local SNR is low ($\leq 2$). For these off-patch
and low-QC in-patch events, we exclude a $\pm4$-sample
neighborhood around each annotated time from the pointwise
loss, avoiding forced positive supervision on locally
ambiguous waveforms.

For positive events, small timing differences across channels
can make single-point labels overly restrictive. We therefore replace hard labels with Gaussian targets centered on the annotated times and truncated to a \(\pm 2\) sample window. This provides graded supervision over
a short temporal neighborhood, accommodating minor timing
variations while still encouraging responses near the
annotated event centers.

We also optimize a masked, class-balanced BCE loss with Gaussian soft
targets and OHEM-selected negatives, supplemented by a positive-event
bag loss to tolerate small temporal offsets:
\begin{equation}
\begin{aligned}
\mathcal{L}
&= \mathcal{L}_{\mathrm{BCE}}
+ \lambda_{\mathrm{bag}}
\frac{\sum_i w_i \log(1+e^{-\widetilde{z}_i})}
{\max(1,\sum_i w_i)},\\
\widetilde{z}_i
&= \tau\log\!\left(
\frac{1}{|\mathcal{B}_i|}
\sum_{t\in\mathcal{B}_i}e^{z_t/\tau}
\right).
\end{aligned}
\end{equation}
Here, $z_t$ is the predicted logit and $\mathcal{B}_i$ contains
unmasked samples within $\pm6$ samples of valid event center $i$.
We set $\tau=0.15$ and $\lambda_{\mathrm{bag}}=0.03$.
The weights $w_i$ combine event SNR and label confidence. These weights also reweight the positive BCE component.

We further introduce score and SNR based postprocessing to
improve detection quality. Candidates reaching the direct-keep
threshold are retained without SNR gating, while lower-scoring
candidates that meet the base detection threshold are retained
only if they satisfy multichannel event-SNR criteria. This
strategy complements detector scores with local waveform
evidence when evaluating uncertain candidates.

\subsection{VanillaCluster}

VanillaCluster augments standard GMM clustering by concatenating relative cross-channel amplitude features with HuiduRep embeddings. For each event, peak-to-peak amplitudes are computed on the four physical channels and normalized by their sum. The resulting amplitude fractions are projected onto orthonormal Helmert contrasts and standardized within each recording.

Cross-fitted waveform templates are constructed from core events with GMM posterior $\geq0.9$, excluding the lowest 25\% of detector scores within each cluster's core pool from template estimation only. Each event is evaluated against templates built from the
opposite fold of alternating 1-s blocks.
A single reassignment among the top-three GMM components $\mathcal K_i$ uses
\[
\hat k_i=\arg\max_{k\in\mathcal K_i}
\left\{\log r_{ik}
+\operatorname{clip}\!\left(2[E_{i,k_i^0}-E_{ik}],-4,4\right)\right\},
\]
where $r_{ik}$ is the GMM posterior, $k_i^0$ is the initial assignment, and $E_{ik}$ is the waveform--template residual minimized over bounded amplitude scaling and small temporal shifts.
This bounded correction incorporates waveform consistency while preserving the initial clustering evidence.

\section{Datasets}
\label{sec:datasets}
We use two datasets in this study. The first consists of real recordings from the mouse cortex provided by the International Brain Laboratory (IBL) \cite{10.7554/eLife.63711}, accompanied by spike sorting results obtained using Kilosort \cite{pachitariu2024kilosort2}. These sorting outputs should be treated as noisy and incomplete pseudo-labels rather than ground truth, as they may contain false positives and omit true spike events. We use 6 IBL recordings for model training and 2 for validation, extracting 1,800 seconds of data from each recording.

The second is the Hybrid Janelia dataset provided through SpikeForest \cite{magland2020spikeforest}, which is constructed by injecting real spikes into synthetic background recordings. This construction provides ground-truth labels for the injected events. It comprises static and drift subsets, with the latter introducing simulated drift in spike waveforms to simulate realistic recording conditions. We evaluate performance on ground-truth units with \(\mathrm{SNR} \geq 3\), consistent with the SNR criterion used by SimSort \cite{zhang2025simsortdatadrivenframeworkspike} and HuiduRep \cite{cao2026huidureprobustselfsupervisedframework}. For evaluation, both static and drift subsets have 9 recordings.

\section{Experiments}
\label{sec:experiments}

\subsection{Implementation Details}
All experiments in this paper were conducted on an NVIDIA RTX PRO 6000 GPU. For VanillaDet, we used a warm-up followed by cosine annealing with warm restarts \cite{loshchilov2017sgdrstochasticgradientdescent}. Early stopping was applied based on the validation area under the precision-recall curve (AUPRC). To prevent gradient explosion, Query-Key normalization \cite{henry2020querykeynormalizationtransformers} was also applied. All hyperparameters were selected using only the IBL split; Hybrid Janelia were used only for evaluation. Code is available at https://github.com/IgarashiAkatuki/VanillaSort

\subsection{Performance Evaluation}
In experiments, we separately evaluated the performance of the detector and that of the complete spike sorting pipeline.

\textbf{VanillaDet Performance.} As shown in Table \ref{tab:detection_results}, despite being trained on noisy pseudo-labels, VanillaDet substantially outperforms threshold-based detectors on both subsets of Hybrid Janelia and consistently surpasses SimSort's detector \cite{zhang2025simsortdatadrivenframeworkspike} across all evaluation metrics. 


\begin{table}[t]
    \centering
    \caption{Spike detection results on Hybrid Janelia.
    Values are mean $\pm$ Standard Error of the Mean (SEM) across recordings.
    The threshold baseline uses the best-performing threshold
    selected by grid search.
    Best results for each metric and subset are in \textbf{bold}.}
    \label{tab:detection_results}
    \vspace{4pt}
    \small
    \renewcommand{\arraystretch}{0.8}
    \setlength{\tabcolsep}{2pt}

    \begin{tabular*}{\columnwidth}{
        @{\extracolsep{\fill}}llccc@{}
    }
        \toprule
        \textbf{Subset}
        & \textbf{Metric}
        & \textbf{Threshold}
        & \textbf{SimSort}
        & \textbf{VanillaDet} \\
        \midrule

        \multirow{3}{*}{Static}
        & Accuracy
        & $0.61 \pm 0.03$
        & $0.72 \pm 0.03$
        & $\mathbf{0.74 \pm 0.02}$ \\

        & Recall
        & $0.71 \pm 0.02$
        & $0.84 \pm 0.02$
        & $\mathbf{0.86 \pm 0.02}$ \\

        & Precision
        & $0.81 \pm 0.02$
        & $0.82 \pm 0.02$
        & $\mathbf{0.85 \pm 0.01}$ \\
        \midrule

        \multirow{3}{*}{Drift}
        & Accuracy
        & $0.60 \pm 0.03$
        & $0.68 \pm 0.03$
        & $\mathbf{0.71 \pm 0.02}$ \\

        & Recall
        & $0.70 \pm 0.03$
        & $0.82 \pm 0.02$
        & $\mathbf{0.83 \pm 0.02}$ \\

        & Precision
        & $0.80 \pm 0.02$
        & $0.81 \pm 0.02$
        & $\mathbf{0.84 \pm 0.01}$ \\
        \bottomrule
    \end{tabular*}
\end{table}

\begin{figure}[t]
    \centering
    \includegraphics[width=0.80\columnwidth]{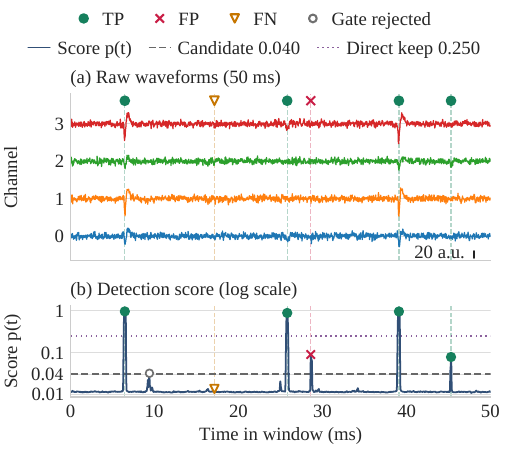}
    \caption{Example VanillaDet detections.
    (a) Raw four-channel waveforms, vertically offset for clarity.
    (b) Sigmoid detection scores on a logarithmic scale.}
    \label{fig:vanilladet_detection}
\end{figure}

\begin{table*}[t]
    \centering
    \small
    \renewcommand{\arraystretch}{0.8}
    \setlength{\tabcolsep}{3pt}

    \caption{Spike sorting results. Values are mean $\pm$ SEM across recordings.
    Results for other methods are obtained from SpikeForest or their original paper.
    Best results in each column are shown in \textbf{bold}. Within each HuiduRep block, components are added cumulatively. Asterisks indicate significant improvements over the corresponding
HuiduRep baseline within each block (Wilcoxon test \cite{wilcoxon1945individual}
on paired recording scores, $p < 0.05$).}
    \label{tab:my_label}

    \vspace{3pt}
    \begin{tabular*}{\textwidth}{
        @{\extracolsep{\fill}}lcccccc@{}
    }
        \toprule
        \multirow{2}{*}{\textbf{Method}}
        & \multicolumn{3}{c}{\textbf{Hybrid\_Janelia-Static}}
        & \multicolumn{3}{c}{\textbf{Hybrid\_Janelia-Drift}} \\
        \cmidrule(lr){2-4}
        \cmidrule(lr){5-7}
        & \textbf{Accuracy}
        & \textbf{Recall}
        & \textbf{Precision}
        & \textbf{Accuracy}
        & \textbf{Recall}
        & \textbf{Precision} \\
        \midrule

        HerdingSpikes2~\cite{HILGEN20172521}
        & $0.35\pm0.01$
        & $0.44\pm0.02$
        & $0.53\pm0.01$
        & $0.29\pm0.01$
        & $0.37\pm0.02$
        & $0.48\pm0.02$ \\

        IronClust~\cite{jun_magland_2025_ironclust}
        & $0.57\pm0.04$
        & $\mathbf{0.81\pm0.01}$
        & $0.60\pm0.04$
        & $0.54\pm0.03$
        & $\mathbf{0.71\pm0.02}$
        & $0.65\pm0.03$ \\

        JRClust~\cite{Jun101030}
        & $0.47\pm0.04$
        & $0.63\pm0.02$
        & $0.59\pm0.03$
        & $0.35\pm0.03$
        & $0.48\pm0.03$
        & $0.57\pm0.02$ \\

        KiloSort~\cite{Pachitariu061481}
        & $0.60\pm0.02$
        & $0.65\pm0.02$
        & $0.72\pm0.02$
        & $0.51\pm0.02$
        & $0.62\pm0.01$
        & $0.72\pm0.03$ \\

        KiloSort2~\cite{pachitariu2024kilosort2}
        & $0.39\pm0.03$
        & $0.37\pm0.03$
        & $0.51\pm0.03$
        & $0.30\pm0.02$
        & $0.31\pm0.02$
        & $0.57\pm0.04$ \\

        KiloSort4~\cite{pachitariu2024spike}
        & $0.40\pm0.03$
        & $0.45\pm0.03$
        & $0.52\pm0.05$
        & $0.34\pm0.02$
        & $0.35\pm0.02$
        & $0.61\pm0.03$ \\

        MountainSort4~\cite{magland2025mountainsort4}
        & $0.59\pm0.02$
        & $0.73\pm0.01$
        & $0.74\pm0.03$
        & $0.36\pm0.02$
        & $0.57\pm0.02$
        & $0.61\pm0.03$ \\

        MountainSort5~\cite{magland2025mountainsort5}
        & $0.40\pm0.06$
        & $0.50\pm0.05$
        & $0.52\pm0.08$
        & $0.33\pm0.04$
        & $0.40\pm0.03$
        & $0.64\pm0.05$ \\

        SpykingCircus~\cite{10.7554/eLife.34518}
        & $0.57\pm0.01$
        & $0.63\pm0.01$
        & $0.75\pm0.03$
        & $0.48\pm0.02$
        & $0.55\pm0.02$
        & $0.68\pm0.03$ \\

        Tridesclous~\cite{pouzat_garcia_2025_tridesclous}
        & $0.54\pm0.03$
        & $0.66\pm0.02$
        & $0.59\pm0.04$
        & $0.37\pm0.02$
        & $0.52\pm0.03$
        & $0.55\pm0.04$ \\

        SimSort~\cite{zhang2025simsortdatadrivenframeworkspike}
        & $0.62\pm0.04$
        & $0.68\pm0.04$
        & $0.77\pm0.03$
        & $0.56\pm0.03$
        & $0.63\pm0.03$
        & $0.69\pm0.03$ \\

        \midrule

        \textbf{HuiduRep without DAE}
        & $0.69\pm0.02$
        & $0.72\pm0.02$
        & $0.87\pm0.01$
        & $0.56\pm0.02$
        & $0.61\pm0.02$
        & $0.83\pm0.01$ \\

        \textbf{+VanillaDet}
        & $0.70\pm0.02^{*}$
        & $0.73\pm0.02^{*}$
        & $0.87\pm0.01$
        & $0.57\pm0.02^{*}$
        & $0.63\pm0.02^{*}$
        & $0.83\pm0.02$ \\

        \textbf{+Amplitude Features}
        & $0.71\pm0.02^{*}$
        & $0.74\pm0.02^{*}$
        & $0.87\pm0.01$
        & $0.59\pm0.02^{*}$
        & $0.63\pm0.02^{*}$
        & $0.84\pm0.02^{*}$ \\

        \textbf{+Guided Reassignment (VanillaSort)}
        & $\mathbf{0.73\pm0.02}^{*}$
        & $0.76\pm0.01^{*}$
        & $\mathbf{0.88\pm0.01}^{*}$
        & $0.61\pm0.02^{*}$
        & $0.65\pm0.02^{*}$
        & $\mathbf{0.85\pm0.02}^{*}$ \\

        \midrule

        \textbf{HuiduRep with DAE}
        & $0.70\pm0.02$
        & $0.75\pm0.02$
        & $0.85\pm0.01$
        & $0.60\pm0.02$
        & $0.65\pm0.02$
        & $0.83\pm0.01$ \\

        \textbf{+VanillaDet}
        & $0.71\pm0.02^{*}$
        & $0.75\pm0.01$
        & $0.86\pm0.01$
        & $0.61\pm0.02^{*}$
        & $0.66\pm0.01^{*}$
        & $0.83\pm0.02$ \\

        \textbf{+Amplitude Features}
        & $0.71\pm0.02^{*}$
        & $0.75\pm0.02$
        & $0.86\pm0.01^{*}$
        & $0.62\pm0.02^{*}$
        & $0.66\pm0.02^{*}$
        & $0.83\pm0.02$ \\

        \textbf{+Guided Reassignment (VanillaSort)}
        & $\mathbf{0.73\pm0.01}^{*}$
        & $0.77\pm0.01^{*}$
        & $0.87\pm0.01^{*}$
        & $\mathbf{0.64\pm0.02}^{*}$
        & $0.68\pm0.01^{*}$
        & $\mathbf{0.85\pm0.01}^{*}$ \\

        \bottomrule
    \end{tabular*}
\end{table*}

\textbf{Case Study.} As shown in Fig. \ref{fig:vanilladet_detection}, within this segment, the gate rejects a candidate exceeding the base threshold of 0.04, while retaining a true positive whose score is below the direct-keep threshold of 0.25. For SNR gating, we require an event to have SNR $\geq3$ on one channel and SNR of $\geq2$ on another one. This illustrates how event-SNR information complements detection scores to selectively filter candidates.

\textbf{VanillaSort Performance.} For spike sorting pipeline, we follow HuiduRep's experimental settings and statistical analysis
\cite{cao2026huidureprobustselfsupervisedframework}. Following HuiduRep, we set $K_{\mathrm{GMM}} \approx \hat{N}_{\mathrm{units}} + 6$, where $\hat{N}_{\mathrm{units}}$ is a biologically informed estimate of the expected local unit count. With ground-truth events matched within $\pm6$ samples, we compute accuracy (Acc.) as
$n_2/(n_1+n_2+n_3)$, precision as $n_2/(n_2+n_3)$, and recall
as $n_2/(n_1+n_2)$, where $n_1$, $n_2$, and $n_3$ count
missed ground-truth events, matched events, and unmatched
predictions, respectively. 

We integrate VanillaDet and VanillaCluster into a spike sorting pipeline to evaluate their contributions to the final neuronal assignment. By using the HuiduRep pipeline as the backbone, we first replace its threshold-based detector with VanillaDet and then additionally replace its original cluster with VanillaCluster, yielding the VanillaSort pipeline.

As shown in Table \ref{tab:my_label}, we compare VanillaSort with several competitive spike sorting methods, including the KiloSort \cite{pachitariu2024kilosort2}, MountainSort \cite{CHUNG20171381} and SimSort \cite{zhang2025simsortdatadrivenframeworkspike}. Compared with the corresponding HuiduRep baselines with or without Denoising Autoencoder (DAE) \cite{bengio2013generalizeddenoisingautoencodersgenerative}, VanillaSort improves accuracy by 3–4 percentage points on static recordings and 4–5 points on drift recordings.

\begin{table}[t]
    \centering
    \caption{Ablation study of VanillaDet on Hybrid Janelia.
    Each factor is varied while the remaining settings are fixed.
    Hard denotes a single-point hard label. Default settings are shown with asterisks. }
    \label{tab:vanilladet_ablation}

    \vspace{4pt}
    \small
    \renewcommand{\arraystretch}{0.80}
    \setlength{\tabcolsep}{2pt}

    \begin{tabular*}{\columnwidth}{
        @{\extracolsep{\fill}}lcccc@{}
    }
        \toprule
        \textbf{Factor}
        & \textbf{Setting}
        & \textbf{Acc.}
        & \textbf{Recall}
        & \textbf{Precision} \\
        \midrule

        \multirow{4}{*}{\shortstack[l]{Bag loss\\weight}}
        & $0$    & $0.716\pm0.012$ & $0.844\pm0.013$ & $0.824\pm0.008$ \\
        & $\mathbf{0.03}^{*}$ & $0.725\pm0.012$ & $0.843\pm0.013$ & $0.840\pm0.008$ \\
        & $0.06$ & $0.719\pm0.012$ & $0.843\pm0.013$ & $0.830\pm0.008$ \\
        & $0.09$ & $0.706\pm0.013$ & $0.842\pm0.013$ & $0.812\pm0.009$\\
        \midrule

        \multirow{4}{*}{\shortstack[l]{Uncertain\\mask}}
        & Off     & $0.665\pm0.011$ & $0.843\pm0.013$ & 
        $0.760\pm0.008$\\
        & $\pm 2$ & $0.616\pm0.008$ & $0.853\pm0.013$ & $0.690\pm0.008$ \\
        & $\mathbf{\pm 4^{*}}$ & $0.725\pm0.012$ & $0.843\pm0.013$ & $0.840\pm0.008$ \\
        & $\pm 8$ & $0.660\pm0.010$ & $0.842\pm0.014$ & $0.754\pm0.008$ \\
        \midrule

        \multirow{4}{*}{\shortstack[l]{Gaussian\\label}}
        & Hard     & $0.704\pm0.013$ & $0.844\pm0.013$ &
        $0.809\pm0.009$\\
        & $\pm 1$  & $0.688\pm0.013$ & $0.843\pm0.013$ &
        $0.788\pm0.009$\\
        & $\mathbf{\pm 2}^{*}$  & $0.725\pm0.012$ & $0.843\pm0.013$ & $0.840\pm0.008$ \\
        & $\pm 4$  & $0.623\pm0.012$ & $0.858\pm0.011$ &
        $0.693\pm0.010$\\
        
        \bottomrule
    \end{tabular*}
\end{table}
\subsection{Ablation Study}

We conducted ablation studies on the bag loss weight, uncertain mask and Gaussian target label, with detailed results reported in Table \ref{tab:vanilladet_ablation}. Results show that removing the
bag loss or uncertainty mask or replacing Gaussian targets
with hard labels, reduces accuracy and precision.
Wider masks and Gaussian targets also reduce accuracy,
indicating that broader temporal tolerance is not
uniformly beneficial.


Table \ref{tab:my_label} shows the gains from sequentially adding
amplitude features and guided reassignment.
The score-informed reassignment stage further improves
accuracy by 2 percentage points on both subsets,
with and without DAE. In this stage, detector scores filter
the GMM core-event pool for cross-fitted template estimation,
extending their role beyond candidate gating to support
downstream neuronal assignment.

\section{Conclusion}
\label{sec:conclusion}

VanillaSort combines visibility-aware detection trained on imperfect pseudo-labels with spatially augmented, template-guided clustering. Experiments on Hybrid Janelia show improved detection and sorting performance under both static and drift conditions.


\section{Acknowledgments}
The authors declare no conflicts of interest.

\section{COMPLIANCE WITH ETHICAL STANDARDS}
\label{sec:ethic}
This study uses only publicly available datasets, which were collected and shared in compliance with institutional and ethical guidelines as stated in the original publications.

\begingroup
\small
\bibliographystyle{IEEEbib}
\bibliography{strings,refs}
\endgroup

\end{document}